\documentclass[RNAAS]{aastex62}

\begin{document}

\title{MN44: a luminous blue variable running away from Westerlund\,1}

\correspondingauthor{V.V. Gvaramadze}
\email{vgvaram@mx.iki.rssi.ru}

\author[0000-0003-1536-8417]{V.V. Gvaramadze}
\affiliation{Sternberg Astronomical Institute, Lomonosov Moscow State University,\\
Universitetskij Pr. 13, Moscow 119992, Russia}


\section{} 

MN44 is a bona fide luminous blue variable (LBV) discovered through the detection of its circumstellar nebula with 
{\it Spitzer} \citep{Gvaramadze2010} and follow up spectroscopic and photometric observations of the central star of 
the nebula \citep{Gvaramadze2015}. Like many other LBVs, MN44 is located in the field and therefore is a runaway star 
\citep{Gvaramadze2012}. With the advent of the {\it Gaia} second data release \citep[DR2;][]{Gaia2018} it become possible 
to derive the space velocity of this star and to search for its parent star cluster. 

Using the {\it Gaia} DR2 proper motion of MN44, $\mu _\alpha \cos \delta= -4.270\pm0.211 \, {\rm mas} \, {\rm yr}^{-1}$ and 
$\mu _\delta=-6.106\pm0.143 \, {\rm mas} \, {\rm yr}^{-1}$, and the Bayesian distance estimate of $d=5.2^{+2.7} _{-1.7}$ 
kpc \citep[based on the {\it Gaia} DR2 data;][]{Bailer2018}, we calculated the peculiar transverse velocity of this star 
$v_{\rm tr}=(v_{\rm l}^2 +v_{\rm b}^2)^{1/2}$, where $v_{\rm l}$ and $v_{\rm b}$ are, respectively, the peculiar velocity 
components along the Galactic longitude and latitude. For this, we adopted the solar Galactocentric distance of $R_0 = 
8.0$ kpc and the circular Galactic rotation velocity of $\Theta _0 =240 \, {\rm km} \, {\rm s}^{-1}$ \citep{Reid2009}, and 
the solar peculiar motion of $(U_{\odot},V_{\odot},W_{\odot})=(11.1,12.2,7.3) \, {\rm km} \, {\rm s}^{-1}$ 
\citep{Schonrich2010}. For the error calculation, only the errors of the proper motion measurements were considered. 
Assuming $d=5.2$ kpc, we found $v_{\rm l}\approx-73\pm4 \, {\rm km} \, {\rm s}^{-1}$, $v_{\rm b}\approx-16\pm5 \, {\rm km} 
\, {\rm s}^{-1}$ and $v_{\rm tr}\approx75\pm4 \, {\rm km} \, {\rm s}^{-1}$, i.e. MN44 is moving away from the Galactic 
plane and its trajectory is inclined to this plane by an angle $\alpha\approx12\degr\pm4\degr$.

Using the WEBDA data base \citep{Mermiliod1995}, we searched for young massive star clusters within the error cone 
of the past trajectory of MN44. Much to our surprise, we found that MN44 is running away from one of the most massive 
Galactic star clusters --- Westerlund\,1 (note that the line connecting the star and the cluster is inclined to the 
Galactic plane at $\approx9\degr$). At the adopted distance, the angular separation between the cluster and the star of 
$\approx4\fdg5$ corresponds to the projected linear separation of $\approx400$ pc, which implies the kinematic age of MN44 
of $t_{\rm kin}\approx5.4$\,Myr. Both the distance to and the kinematic age of MN44 are in good agreement with the distance 
and age estimates for Westerlund\,1 of 4--5 kpc and $\sim5$ Myr, respectively \citep[e.g.][and references therein]{Fenech2018}.
This strongly suggests that Westerlund\,1 is the parent cluster to MN44. An indirect support to this possibility comes from 
the almost identical visual extinction towards both objects of 10.8 mag \citep{Negueruela2010} and 10.2 mag 
\citep{Gvaramadze2015}, respectively. 

\begin{figure}[h!]
 \begin{center}
 \includegraphics[scale=0.8,angle=0]{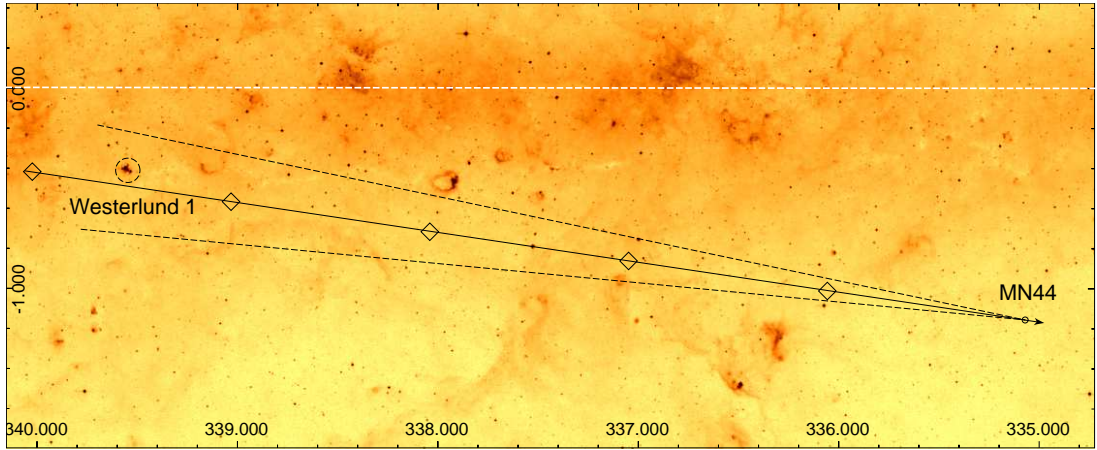}
 \caption{{\it Midcourse Space Experiment} satellite \citep{Price2001} 8.3 \micron \, image of the field containing 
 Westerlund\,1 (indicated by a dashed circle) and MN44 (marked by a circle). The arrow shows the direction of motion 
 of MN44, while a solid line indicates the trajectory of MN44 (with 1$\sigma$ uncertainties shown by dashed lines) for 
 $d=4$\,kpc. The positions of MN44 1, 2, 3, 4 and 5 Myr ago are marked by diamonds. The image is oriented with Galactic 
 longitude (in units of degrees) increasing to the left and Galactic latitude increasing upwards. The Galactic plane is 
 shown by a white dashed line. At the distance of 4 kpc, 1\degr corresponds to $\approx68.8$ pc. \label{fig:wd1}}
 \end{center}
 \end{figure}

We also calculated velocity for $d=4$ kpc to account for the possibility that both the star and the cluster are located 
at a shorter distance. We found $v_{\rm l}=-73\pm3 \, {\rm km} \, {\rm s}^{-1}$, $v_{\rm b}=-11\pm4 \, {\rm km} \, 
{\rm s}^{-1}$ and $v_{\rm tr}=74\pm3 \, {\rm km} \, {\rm s}^{-1}$. These velocities imply $\alpha=8\fdg5\pm3\fdg0$ and 
$t_{\rm kin}\approx4.2$ Myr. At this distance, the trajectory of MN44 almost exactly overlaps with the line connecting the 
star and the cluster (see Fig.\,\ref{fig:wd1}).

Both estimates of $t_{\rm kin}$ indicate that MN44 was ejected in less than 1\,Myr after the cluster formation, implying 
that Westerlund\,1 was mass-segregated and collisional at birth \citep[cf.][]{Fujii2012} and that many other massive stars 
should run away from this cluster \citep[cf.][]{Gvaramadze2008}. If final {\it Gaia} parallaxes will confirm that MN44 and 
Westerlund\,1 are located at the same distance, then MN44 would represent the first star ejected from this cluster, whose 
runaway status was derived on the basis of proper motion measurements \citep[cf.][]{Clark2014}.

This work is supported by the Russian Foundation for Basic Research grant 16-02-00148.

\end{document}